\documentstyle[12pt]{article}
\begin{document}

\bigskip

\centerline {\bf {Origin of Radially Increasing Stellar Scaleheight}}
\centerline {\bf {in a Galactic Disk}}
\medskip

\bigskip

\bigskip

\centerline{Chaitra A. Narayan and Chanda J. Jog}
\centerline {Department of Physics, Indian Institute of Science}
\centerline { Bangalore 560 012, INDIA.}
\centerline {email: chaitra@physics.iisc.ernet.in,cjjog@physics.iisc.ernet.in}

\bigskip

\bigskip

\centerline{Astron. \& Astrophys. Letters,  2002, in press.}

\newpage
\noindent {\bf Abstract} 

\noindent For the past twenty years, it has been accepted that the vertical 
scaleheight of the stellar disk in spiral galaxies is constant with radius.
However, there is no clear physical explanation for this in the literature.
Here we calculate the vertical stellar scaleheight for a self-gravitating stellar disk 
including the additional gravitational force of the HI and H$_2$ gas and the dark 
matter halo.
We apply our model to two edge-on galaxies, NGC 891 and NGC 
4565, and find that the resulting scaleheight shows a linear increase of 
nearly a factor of two within the optical disk for both these galaxies.
Interestingly, we show that the observed data when looked at closely,
do not imply a constant scaleheight but actually support this moderate 
flaring in scaleheight.

\noindent Running Title : Radially Increasing Stellar Scaleheight

\bigskip


\newpage

\noindent {\bf 1. Introdution}

\noindent In the pioneering studies on vertical luminosity distribution of edge-on 
stellar disks, van der Kruit and Searle (1981a,b) found that the vertical 
scaleheight of stars is independent of radius in NGC 4244, NGC 5907, NGC
4565 and NGC 891.
This was later confirmed for NGC 4244 by Fry et.al. (1999) and for NGC 891
by Kylafis \& Bahcall (1987) and also for other galaxies. In fact, the constant 
stellar scaleheight has become a well-established paradigm.
Recent studies, however, have shown (Kent, Dame \& Fazio 1991 and de Grijs \&
Peletier 1997) a moderate increase with radius in stellar scaleheight.

The constancy of scaleheight comes about when the scalelength 
with which the vertical velocity dispersion falls off exponentially along the radius, 
$h_{\mathrm{vel}}$, is twice the radial density scalelength, $h_{\mathrm{R}}$, in 
a galactic disk (van der Kruit and Searle 1981a).
However, this relation has no theoretical basis and is used only to obtain the flat
scaleheight.
We note that if the velocity falls off slower than the above rate 
($h_{\mathrm{vel}} > 2 h_{\mathrm{R}}$), then it results in a flaring disk 
whereas the converse ($h_{\mathrm{vel}} < 2 h_{\mathrm{R}}$) gives rise to a disk 
which tapers down in thickness.
These results are valid when the density distribution is defined only by the 
gravitational force of the stellar disk.

In reality, the molecular and atomic hydrogen gas and the dark matter halo
exert a considerable gravitational force on the stellar disk and hence 
should be included in defining the stellar scaleheight.
The effect of these components gains importance at large galactocentric 
radii, where the surface densities of all the components are comparable 
to each other, whereas in the inner galactic disk the stellar component
dominates over the others.
We show that it is because of this additional gravitational force, that 
the stellar scaleheight is not exponentially flaring at large radii but 
instead it shows a controlled linear variation.
Earlier we have proposed (Narayan \& Jog 2002) a gravitationally 
coupled, three-component disk plus halo model which predicts the scaleheight 
curves of HI, H$_2$ and the stellar disk of our Galaxy, which are shown to be 
in good agreement with observations.

In this Letter, we apply the same model for two external galaxies, NGC 891 
and NGC 4565, and compare the stellar scaleheights obtained with observations 
available in the literature. These galaxies are chosen because these are two 
of the prototypical edge-on galaxies from the sample of van der Kruit \&
Searle (1981a,b) which also have detailed HI and H$_2$ surface 
density values from observations.

In Sect. 2, we discuss the details of the model used.
The resulting scaleheights for NGC 891 and NGC 4565 are compared with 
observations in Sect. 3. 
Section 4 contains discussion and Sect. 5 gives a summary of the 
conclusions. 
 
\noindent {\bf {2. Calculation for vertical stellar scaleheight}}

\noindent The treatment for obtaining the vertical stellar scaleheight is based
on a three-component disk plus halo model, as described by Narayan \&
Jog (2002), which is briefly summarized below.

The force equation along the $z$-axis or the equation of hydrostatic
equilibrium is given by (e.g., Rolhfs 1977):

$$ \mathrm{\frac {\langle{(\mathit{v}_z)^2_i} \rangle}{\rho_i} \: 
   \frac {d {\rho_i}}{d{\mathit{z}}} \: = \: {(\mathit{K}_z)_s} \: + 
   \: {(\mathit{K}_z)_{HI}} \: + \: {(\mathit{K}_z)_{H_2}} 
   \: + \: {(\mathit{K}_z)_{DM}} }\eqno (1) $$

\noindent where $\rho$ is the mass density, 
$(K_{\mathrm{z}}) \: = \: - {{\partial}{\psi}}/ {{\partial}z} $
is the force per unit mass along the $z$-axis, $\psi$ is the corresponding
potential, and the subscript $i$ = s, HI, H$_2$ and DM denotes 
these quantities for stars, HI, H$_2$ and the dark matter halo respectively.
We take the root mean square of
the vertical velocities of a component $\langle{(v_{\mathrm{z}})^2_{\mathrm{i}}
\rangle ^ \frac {1}{2}}$
or the random velocity dispersion at a radius $R$ and treat the component as
being isothermal along $z$. 
The right hand side of equation (1) gives the total vertical force due to all 
the components.

\noindent For a thin disk, the joint Poisson equation reduces to :

$$\mathrm{ \frac {d^2{\psi_s}}{d{\mathit{z}^2}} \: + \frac {d^2{\psi_{HI}}}
{d{\mathit{z}^2}} \: + \frac {d^2{\psi_{H_2}}}{d{\mathit{z}^2}} }\: = \:  4 \pi G \left 
({\rho}_{\mathrm{s}} \: +  \: {\rho}_{\mathrm{HI}} \: + \: {\rho}_{\mathrm{H_2}} 
\right ) \eqno (2) $$

The coupled equations (1) and (2) are solved for the density $\rho$ 
versus $z$, as a boundary
value problem using standard numerical techniques (Press et.al. 1986). 
To obtain the vertical density distribution of a component, we need to 
know the surface density of the component which is used as a boundary 
condition, and its vertical velocity dispersion.
This results in a sech$^2$-like density profile, and we use its HWHM to 
define the scaleheight.
We have used a single stellar component for simplicity, though recent 
observations show (de Grijs, Peletier \& van der Kruit 1997)
a more sharply peaked luminosity profile which may be explained as a sum of
several disk components. 

\noindent {\bf {3. Results}}

\noindent {\bf {3.1. NGC 891}}

\noindent {\bf {3.1.1. Scaleheights from our model}}

\noindent NGC 891 is an edge-on spiral ($i = 89^{\mathrm{o}}$) of Hubble type Sb.
It has a very large stellar disk with $h_{\mathrm{R}} = 4.9$ kpc
 (van der Kruit \& Searle 1981b). 
We use a central surface density of 700 M$_{\odot}$ pc$^{-2}$ 
which is the value given for our Galaxy (Binney \& Tremaine 1987),
for the sake of simplicity and also
because the results from this section do not depend 
sensitively on the actual choice of the value used.

Bottema (1993) has measured the vertical velocity dispersions for a set of 
12 spirals and finds that the central velocity dispersion is about 
60 kms$^{-1}$ for an Sb type galaxy, which is what we use for NGC 891 
and NGC 4565. 
The dispersion falls off exponentially and is generally modeled (Lewis \& 
Freeman 1989, Bottema 1993) with a scalelength, $h_{\mathrm{vel}} 
= 2 h_{\mathrm{R}}$ in accordance 
with the van der Kruit \& Searle (1981a) model - see Sect. 1. 
A unique value of $h_{\mathrm{vel}}$ cannot be derived since there are 
only three data points for NGC 891. 
Instead, in our work the ratio $h_{\mathrm{vel}}/h_{\mathrm{R}}$ is allowed to
take a range of values between 1-4. 

The observed surface densities of HI and H$_2$ are taken from Rupen 
(1991) and Sofue \& Nakai (1993) respectively. 
The gas velocity dispersions in the Galaxy are 8 kms$^{-1}$ (Spitzer 1978) and 5 
kms$^{-1}$ (Stark 1984 and Clemens 1985), for HI and H$_2$ respectively.
These are taken to be the same for all the spiral galaxies, as observed 
for HI (Lewis 1984) and H$_2$ (Wilson \& Scoville 1990) and as argued 
theoretically for H$_2$ (Jog \& Ostriker 1988).
Finally, we calculate the gravitational force of the dark matter halo from a 
standard model of Mera, Chabrier \& Schaeffer (1998) with a total halo mass of 
$\sim$ 10$^{12}$ M$_{\odot}$.
These input parameters are used to solve the coupled equations (1) and (2).

Figure 1 shows the scaleheights obtained by the method discussed in Sect. 2
for the ratio $h_{\mathrm{vel}}/h_{\mathrm{R}}$ = 1, 2, 2.5, 3 and 4.
This figure shows that in general, the scaleheight is not flat as deduced by 
van der Kruit \& Searle (1981b).
The ratio of less than 2 can be ruled out since it gives unphysically tapering
scaleheight.

\noindent {\bf {3.1.2. Scaleheights deduced from observed data}}

\noindent Van der Kruit and Searle (1981b) fit their data using an isothermal 
stellar disk model with the following space-luminosity density profile
as applicable for an edge-on system:
$$ \mu(R,z) \: = \: \mu(0,0) \: \left ( \frac{R}{h_{\mathrm{R}}} 
\right )\: K_1 \left ( \frac{R}{h_{\mathrm{R}}} \right ) {\mathrm{sech^2}}
\left ( \frac{z}{z_{\mathrm{o}}} \right )  \eqno{(3)} $$
where  $\mu(0,0)$ is the space-luminosity density at the centre of the 
galaxy, $z_{\mathrm{o}}$ represents the scaleheight and $K_1$ is the modified 
Bessel function of the second kind.
We note that, by definition, the
scaleheight we obtain (see Sect. 2) is not the same as $z_{\mathrm{o}}$ but the 
two are equivalent.
This model gives a single and well-defined sech$^2$ curve on a composite 
$z$-profile, when the $z_{\mathrm{o}}$ is kept constant.  
Figure 2 shows the composite $z$-profile for NGC 891. 
The solid line results when $z_{\mathrm{o}} = 1$ kpc is kept constant with radius.
However, we note that the observed data are spread over an interval about this 
single curve.
We show that a similar spread in the calculated values of surface brightness is 
obtained when $z_{\mathrm{o}}$ is allowed to increase linearly from 0.75 kpc to 
1.25 kpc over 4 disk scalelengths or the entire optical disk.
This range of $z_{\mathrm{o}}$ is found by trial and error such that the model 
values match well with observations.
{\bf{This shows that the spread in the observed data allows for
a linear increase of a factor of 1.7 in}} ${\bf{z_{\mathrm{o}}}}$.
This increase allows us to constrain the  value of $h_{\mathrm{vel}}/
h_{\mathrm{R}}$ to be between 2-2.5 (see Fig. 1).

\noindent {\bf {3.2. NGC 4565}}

\noindent NGC 4565 is an Sb type spiral galaxy with an inclination of 86$^{\mathrm{o}}$.
Figure 3 shows the stellar scaleheight obtained for NGC 4565 using the method 
described in Sect. 2.
The input parameters used here are as follows.
The radial scalelength, $h_{\mathrm{R}}$ is 5.5 kpc (van der Kruit \& Searle 1981a).
The central surface density of stars is taken to be 700 M$_{\odot}$ 
pc$^{-2}$ and the central vertical velocity dispersion = 60 kms$^{-1}$.
The choice of above two numbers is justified in Sect. 3.1.1.
Surface densities of HI and H$_2$ are taken from Rupen (1991) and Sofue
\& Nakai (1994) respectively.
Since the observations for the stellar velocity dispersion are not available for NGC
4565, we vary the ratio $h_{\mathrm{vel}}/h_{\mathrm{R}}$  within the range of 1-4.

Using the composite $z$-profiles (see Sect. 3.1.2), the stellar 
scaleheight for NGC 4565 was shown to be constant at $z_{\mathrm{o}} = 0.79$ kpc 
(van der Kruit \& Searle 1981a).
This was later confirmed by Naslund \& Jorsater (1997).
As in the case of NGC 891, the observed surface brightness values do not 
give rise to a sharp and single sech$^2$ curve. 
Figure 4 shows the numerically obtained composite $z$-profile, and also the 
observed data.
The noted spread in observed data can be reproduced by increasing the 
$z_{\mathrm{o}}$ linearly from 0.4 kpc to 1.0 kpc within the optical disk 
($\leq$ 22 kpc). 
This shows that the observed data of NGC 4565 can hide as much as a factor of 
2.5 variation in scaleheight. This allows us to constrain 
$h_{\mathrm{vel}}/h_{\mathrm{R}}$ to be between 2.5-3 (see Fig. 3).

\noindent  {\bf {4. Discussion}}

\noindent {\bf{1}}. The results in Sect. 3 show that the observational data can 
accommodate a substantial radial variation in scaleheight. However, we would like to
caution that the observed spread in data need not be 
entirely due to variation in scaleheight because, observational uncertainties 
such as the inclination, dust extinction and large measurement
errors in photographic plates could also contribute to it. Morrison 
et al. (1997) conclude from a more complete analysis of the 2-D data that 
the scaleheight cannot be assigned a unique value, instead it lies within 
a range. This supports our model.

\noindent {\bf 2}. From a recent study of 48 edge-on spirals, de Grijs \& Peletier 
(1997) find that the galaxies show a small radial increase in scaleheight in 
agreement with our model.
They also find that the increase in scaleheight is smaller for later-type
galaxies, and have explained this by using the presence of a thick
disk.  We note that, alternatively, this trend can be explained by our model.
This is because there is an increase in the gas mass fraction (Binney \& 
Merrifield 1998) and the dark matter halo fraction (Broeils 1992) for later 
Hubble types. 
Therefore, their confining effect would be higher and
hence the scaleheight increase would be smaller. 

\noindent {\bf {5. Conclusions}}

\noindent We examine the well-established constancy of vertical stellar scaleheight in 
spiral galaxies, and show that this result is not justified physically nor is 
it supported by observations. We show for NGC 891 and NGC 4565 that the 
observational uncertainties can 
hide as much as a factor of 2 increase in the scaleheight within the optical
disk. We show that this variation can be explained theoretically by our model 
which takes account of the gravitational coupling between stars, gas, and the 
dark matter halo.
Such moderate increase in scaleheight is likely to be common in other spiral
galaxies as well.

\bigskip

\noindent {\it {Acknowledgements}}
We would like to thank the referee, R. de Grijs, for his detailed and constructive 
comments which have greatly improved the paper.
 
\bigskip

\noindent {\bf {References}}

\noindent Binney, J., \& Tremaine, S. 1987, Galactic Dynamics

(Princeton: Princeton Univ. Press)

\noindent Binney, J. \& Merrifield, M. 1998, Galactic Astronomy 

(Princeton: Princeton Univ. Press)

\noindent Bottema, R. 1993, A\&A, 275, 16.

\noindent Broeils, A.H. 1992, Ph.D. thesis, Groningen Univ.

\noindent Clemens, D.P. 1985, ApJ, 295, 422.

\noindent de Grijs, R. \& Peletier, R.F. 1997, A\&A, 320, L21.

\noindent de Grijs, R., Peletier, R.F., \& van der Kruit, P.C. 1997,

A \& A, 327, 966

\noindent Fry A.M., Morrison, H.L., Harding, P., \& Boroson, T.A. 

1999, AJ, 118, 1209.

\noindent Jog, C.J. \& Ostriker, J.P. 1988, ApJ, 328, 404.

\noindent Kent, S.M., Dame, T. \& Fazio, G. 1991, ApJ, 378, 131. 

\noindent Kylafis, N.D., \& Bahcall, J.N. 1987, ApJ, 317, 637.

\noindent Lewis, B.M. 1984 ApJ, 285, L453.

\noindent Lewis, J.R. \& Freeman, K.C. 1989, AJ, 97, 139.

\noindent Mera, D., Chabrier, G., \& Schaeffer, R. 1998, A\&A, 330, 

953

\noindent Morrison, H.L., Miller, E.D., Harding, P., Stinebring, D.R.

\& Boroson, T.A. 1997, AJ, 113, 206.

\noindent Narayan, C.A. \& Jog, C.J. 2002, Submitted to A\&A.

\noindent Naslund, M. \& Jorsater, S. 1997, A\&A, 325, 915.

\noindent Press, W.H., Flannery, B.P., Teukolsky, S.A., \& 

Vetterling, W.T. 1986, Numerical Recipes (Cambridge: 

Cambridge Univ. Press), chap. 6.

\noindent Rohlfs, K. 1977, Lectures on Density Wave Theory, (Berlin 

: Springer-Verlag).

\noindent Rupen, M.P. 1991, AJ, 103, 48.

\noindent Sofue, Y. \& Nakai, N. 1993, PASJ 45, 139.

\noindent Sofue, Y. \& Nakai, N. 1994, PASJ 46, 147.

\noindent Spitzer, L. 1978, Physical Processes in the Interstellar 

Medium (New York: John Wiley)

\noindent Stark, A.A. 1984, ApJ, 281, 624

\noindent van der Kruit, P.C. \& Searle, L. 1981a, A\&A, 95, 105.

\noindent van der Kruit, P.C. \& Searle L. 1981b, A\&A, 95, 116.

\noindent Wilson, C.D. \& Scoville, 1990, ApJ, 363, 435.

\begin{center}
{\bf Figure Legends}
\end{center}

\noindent {\bf Figure 1.} Variation in resulting stellar
vertical scaleheight with galactocentric radius for NGC 891.
The variation is found to crucially depend on the ratio 
$h_{\mathrm{vel}}/h_{\mathrm{R}}$.
While the scaleheight is found to decrease by a factor of 50
and 1.1 for a ratio of 1 
and 2 respectively, and increase by a factor of 1.8, 2.8 and 4.5 for the ratio of 
2.5, 3 and 4 respectively, within the optical disk ($\leq$ 20 kpc).
Thus the scaleheight need not be constant with radius for the galaxy.

\noindent {\bf Figure 2.} The plot of surface brightness, $\mu$, versus
the distance from the midplane, $z$, for NGC 891.
This composite $z$-profile is obtained by vertically shifting the 
individual $z$-profiles into coincidence at $z = 1.5$ kpc.
The solid line results when $z_{\mathrm{o}} = 1$ kpc and is independent of 
radius.
When $z_{\mathrm{o}}$ is linearly increased from 0.75-1.25 kpc within 20 kpc, 
surface brightness values (points) disperse over an interval of 1 
mag/arcsec$^2$ at $z = 0$, as is exactly seen in the original data 
(van der Kruit \& Searle 1981b).
The vertical bars indicate the range of the data.
This shows that the data can allow for as much as a factor of 1.7 increase in 
scaleheight.
The deviation of the observed data at high $z$ is due to the thick disk which is
not included in the model.

\noindent {\bf Figure 3.} Variation in resulting stellar
vertical scaleheight with radius for NGC 4565.
The scaleheight is found to decrease by a factor of 50 and 1.1 
for a ratio of 1
and 2 respectively, and increase by a factor of 1.8, 2.7 and 4.4 for the ratio of 
2.5, 3 and 4 respectively, within the optical disk ($\leq$ 22 kpc).
Thus the scaleheight need not be constant with radius for this galaxy too.

\noindent {\bf Figure 4.} The plot of surface brightness, $\mu$, versus
the distance from the midplane, $z$, for NGC 4565, with details similar to Figure 2.
The linear increase in $z_{\mathrm{o}}$ from 0.4-1.0 kpc or an increase of a 
factor of 2.5 
in the scaleheight can be easily hidden by the spread in the observed data.

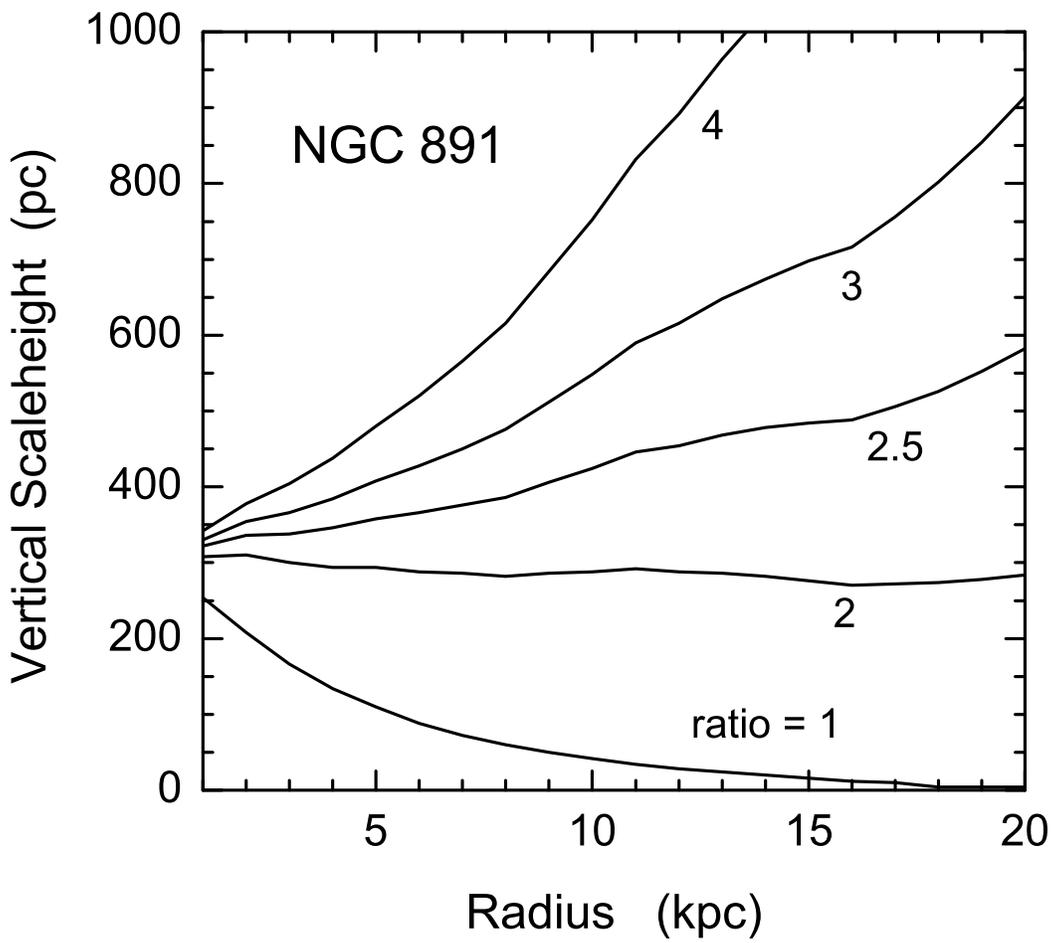
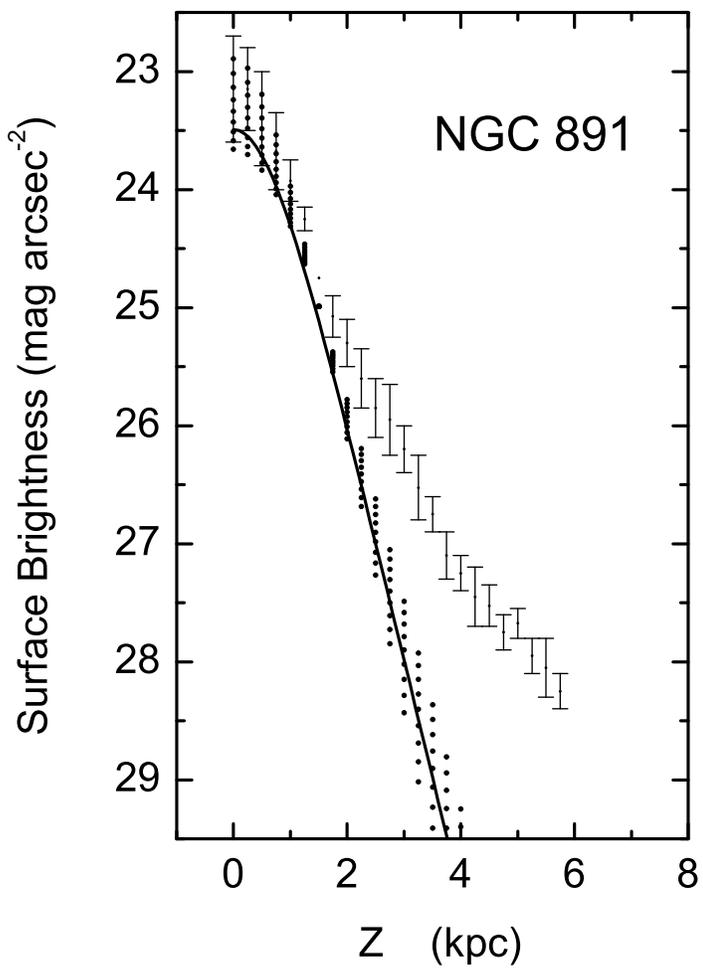
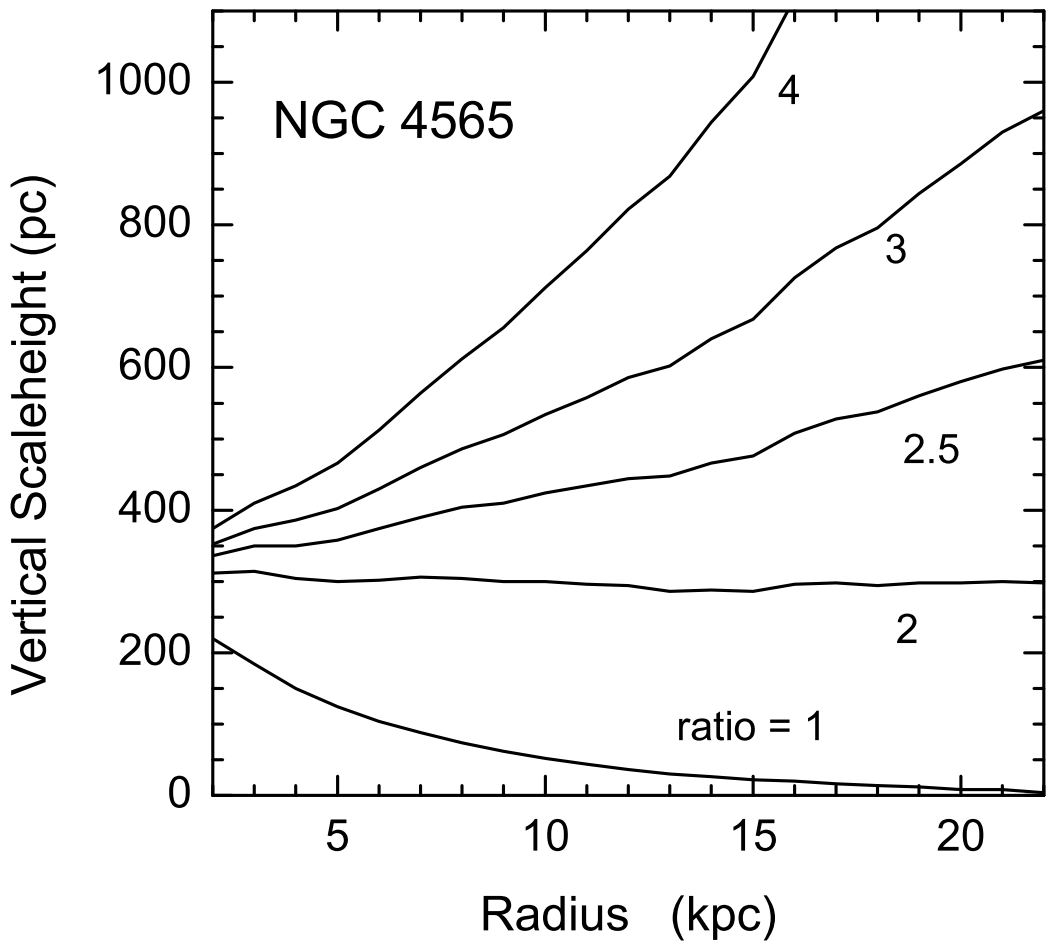
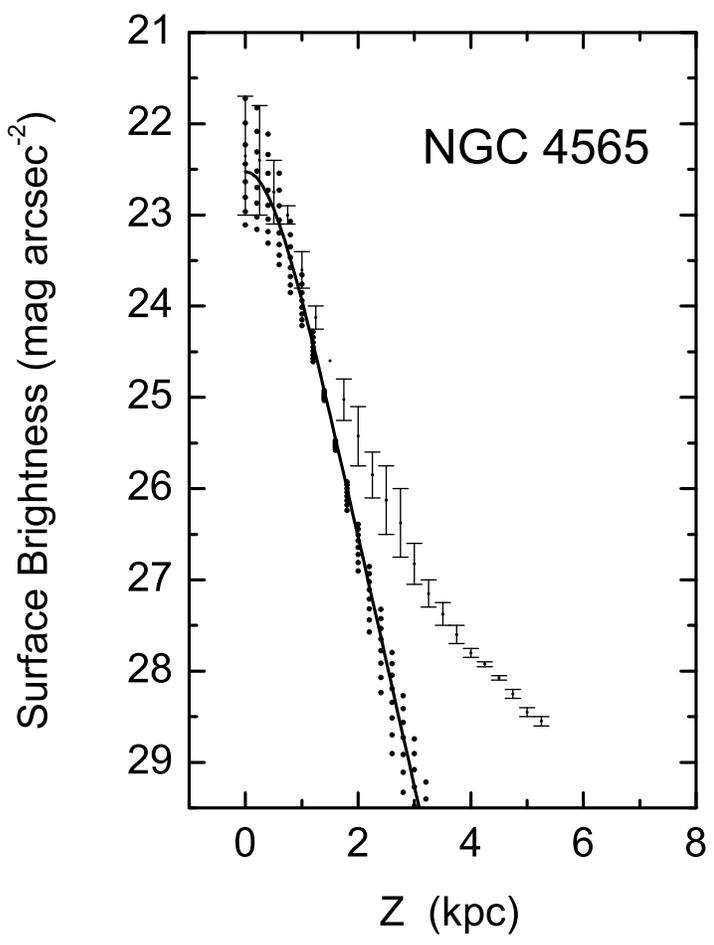
\end{document}